\documentclass[reprint,aps,prd,twocolumn,superscriptaddress,
               amsmath,amssymb,floatfix]{revtex4-1}

\usepackage{graphicx}
\usepackage[mathlines]{lineno}
\usepackage{xspace} 

\newcommand{\eg}{{\it e.g.}\xspace}
\newcommand{\etal}{{\it et al.}\xspace}
\newcommand{\gev}{GeV/c$^2$\xspace}
\newcommand{\tev}{TeV/c$^2$\xspace}
\newcommand{\ba}{$^{133}$Ba\xspace}
\newcommand{\cf}{$^{252}$Cf\xspace}

\begin{document}
\title{Silicon detector results from the first five-tower run of CDMS~II}

\affiliation{Division of Physics, Mathematics \& Astronomy, California Institute of Technology, Pasadena, CA 91125, USA} 
\affiliation{Fermi National Accelerator Laboratory, Batavia, IL 60510, USA}
\affiliation{Lawrence Berkeley National Laboratory, Berkeley, CA 94720, USA}
\affiliation{Department of Physics, Massachusetts Institute of Technology, Cambridge, MA 02139, USA}
\affiliation{Pacific Northwest National Laboratory, Richland, WA 99352, USA}
\affiliation{Department of Physics, Queen's University, Kingston, ON K7L 3N6, Canada}
\affiliation{Department of Physics, Santa Clara University, Santa Clara, CA 95053, USA}
\affiliation{SLAC National Accelerator Laboratory / Kavli Institute for Particle Astrophysics and Cosmology, 2575 Sand Hill Road, Menlo Park 94025, CA}
\affiliation{Department of Physics, Southern Methodist University, Dallas, TX 75275, USA}
\affiliation{Department of Physics, Stanford University, Stanford, CA 94305, USA}
\affiliation{Department of Physics, Syracuse University, Syracuse, NY 13244, USA}
\affiliation{Department of Physics, Texas A \& M University, College Station, TX 77843, USA}
\affiliation{Departamento de F\'{\i}sica Te\'orica and Instituto de F\'{\i}sica Te\'orica UAM/CSIC, Universidad Aut\'onoma de Madrid, 28049 Madrid, Spain}
\affiliation{Department of Physics, University of California, Berkeley, CA 94720, USA}
\affiliation{Department of Physics, University of California, Santa Barbara, CA 93106, USA}
\affiliation{Department of Physics, University of Colorado Denver, Denver, CO 80217, USA}
\affiliation{Department of Physics, University of Evansville, Evansville, IN 47722, USA}
\affiliation{Department of Physics, University of Florida, Gainesville, FL 32611, USA}
\affiliation{School of Physics \& Astronomy, University of Minnesota, Minneapolis, MN 55455, USA}
\affiliation{Physics Institute, University of Z\"{u}rich, Winterthurerstrasse 190, CH-8057, Switzerland}

\author{R.~Agnese} \affiliation{Department of Physics, University of Florida, Gainesville, FL 32611, USA}
\author{Z.~Ahmed} \affiliation{Division of Physics, Mathematics \& Astronomy, California Institute of Technology, Pasadena, CA 91125, USA}
\author{A.J.~Anderson} \affiliation{Department of Physics, Massachusetts Institute of Technology, Cambridge, MA 02139, USA}
\author{S.~Arrenberg} \affiliation{Physics Institute, University of Z\"{u}rich, Winterthurerstrasse 190, CH-8057, Switzerland}
\author{D.~Balakishiyeva} \affiliation{Department of Physics, University of Florida, Gainesville, FL 32611, USA}
\author{R.~Basu~Thakur~} \affiliation{Fermi National Accelerator Laboratory, Batavia, IL 60510, USA}
\author{D.A.~Bauer} \affiliation{Fermi National Accelerator Laboratory, Batavia, IL 60510, USA}
\author{A.~Borgland} \affiliation{SLAC National Accelerator Laboratory / Kavli Institute for Particle Astrophysics and Cosmology, 2575 Sand Hill Road, Menlo Park 94025, CA}
\author{D.~Brandt} \affiliation{SLAC National Accelerator Laboratory / Kavli Institute for Particle Astrophysics and Cosmology, 2575 Sand Hill Road, Menlo Park 94025, CA}
\author{P.L.~Brink} \affiliation{SLAC National Accelerator Laboratory / Kavli Institute for Particle Astrophysics and Cosmology, 2575 Sand Hill Road, Menlo Park 94025, CA}
\author{T.~Bruch} \affiliation{Physics Institute, University of Z\"{u}rich, Winterthurerstrasse 190, CH-8057, Switzerland}
\author{R.~Bunker} \affiliation{Department of Physics, Syracuse University, Syracuse, NY 13244, USA}
\author{B.~Cabrera} \affiliation{Department of Physics, Stanford University, Stanford, CA 94305, USA}
\author{D.O.~Caldwell} \affiliation{Department of Physics, University of California, Santa Barbara, CA 93106, USA}
\author{D.G.~Cerdeno} \affiliation{Departamento de F\'{\i}sica Te\'orica and Instituto de F\'{\i}sica Te\'orica UAM/CSIC, Universidad Aut\'onoma de Madrid, 28049 Madrid, Spain}
\author{H.~Chagani} \affiliation{School of Physics \& Astronomy, University of Minnesota, Minneapolis, MN 55455, USA}
\author{J.~Cooley} \affiliation{Department of Physics, Southern Methodist University, Dallas, TX 75275, USA}
\author{B.~Cornell} \affiliation{Division of Physics, Mathematics \& Astronomy, California Institute of Technology, Pasadena, CA 91125, USA}
\author{C.H.~Crewdson} \affiliation{Department of Physics, Queen's University, Kingston, ON K7L 3N6, Canada}
\author{P.~Cushman} \affiliation{School of Physics \& Astronomy, University of Minnesota, Minneapolis, MN 55455, USA}
\author{M.~Daal} \affiliation{Department of Physics, University of California, Berkeley, CA 94720, USA}
\author{F.~Dejongh} \affiliation{Fermi National Accelerator Laboratory, Batavia, IL 60510, USA}
\author{P.C.F.~Di~Stefano} \affiliation{Department of Physics, Queen's University, Kingston, ON K7L 3N6, Canada}
\author{E.~do~Couto~e~Silva} \affiliation{SLAC National Accelerator Laboratory / Kavli Institute for Particle Astrophysics and Cosmology, 2575 Sand Hill Road, Menlo Park 94025, CA}
\author{T.~Doughty} \affiliation{Department of Physics, University of California, Berkeley, CA 94720, USA}
\author{L.~Esteban} \affiliation{Departamento de F\'{\i}sica Te\'orica and Instituto de F\'{\i}sica Te\'orica UAM/CSIC, Universidad Aut\'onoma de Madrid, 28049 Madrid, Spain}
\author{S.~Fallows} \affiliation{School of Physics \& Astronomy, University of Minnesota, Minneapolis, MN 55455, USA}
\author{E.~Figueroa-Feliciano} \affiliation{Department of Physics, Massachusetts Institute of Technology, Cambridge, MA 02139, USA}
\author{J.~Filippini} \affiliation{Division of Physics, Mathematics \& Astronomy, California Institute of Technology, Pasadena, CA 91125, USA}
\author{J.~Fox} \affiliation{Department of Physics, Queen's University, Kingston, ON K7L 3N6, Canada}
\author{M.~Fritts} \affiliation{School of Physics \& Astronomy, University of Minnesota, Minneapolis, MN 55455, USA}
\author{G.L.~Godfrey} \affiliation{SLAC National Accelerator Laboratory / Kavli Institute for Particle Astrophysics and Cosmology, 2575 Sand Hill Road, Menlo Park 94025, CA}
\author{S.R.~Golwala} \affiliation{Division of Physics, Mathematics \& Astronomy, California Institute of Technology, Pasadena, CA 91125, USA}
\author{J.~Hall} \affiliation{Pacific Northwest National Laboratory, Richland, WA 99352, USA}
\author{R.H.~Harris} \affiliation{Department of Physics, Texas A \& M University, College Station, TX 77843, USA}
\author{S.A.~Hertel} \affiliation{Department of Physics, Massachusetts Institute of Technology, Cambridge, MA 02139, USA}
\author{T.~Hofer} \affiliation{School of Physics \& Astronomy, University of Minnesota, Minneapolis, MN 55455, USA}
\author{D.~Holmgren} \affiliation{Fermi National Accelerator Laboratory, Batavia, IL 60510, USA}
\author{L.~Hsu} \affiliation{Fermi National Accelerator Laboratory, Batavia, IL 60510, USA}
\author{M.E.~Huber} \affiliation{Department of Physics, University of Colorado Denver, Denver, CO 80217, USA}
\author{A.~Jastram} \affiliation{Department of Physics, Texas A \& M University, College Station, TX 77843, USA}
\author{O.~Kamaev} \affiliation{Department of Physics, Queen's University, Kingston, ON K7L 3N6, Canada}
\author{B.~Kara} \affiliation{Department of Physics, Southern Methodist University, Dallas, TX 75275, USA}
\author{M.H.~Kelsey} \affiliation{SLAC National Accelerator Laboratory / Kavli Institute for Particle Astrophysics and Cosmology, 2575 Sand Hill Road, Menlo Park 94025, CA}
\author{A.~Kennedy} \affiliation{School of Physics \& Astronomy, University of Minnesota, Minneapolis, MN 55455, USA}
\author{P.~Kim} \affiliation{SLAC National Accelerator Laboratory / Kavli Institute for Particle Astrophysics and Cosmology, 2575 Sand Hill Road, Menlo Park 94025, CA}
\author{M.~Kiveni} \affiliation{Department of Physics, Syracuse University, Syracuse, NY 13244, USA}
\author{K.~Koch} \affiliation{School of Physics \& Astronomy, University of Minnesota, Minneapolis, MN 55455, USA}
\author{M.~Kos} \affiliation{Department of Physics, Syracuse University, Syracuse, NY 13244, USA}
\author{S.W.~Leman} \affiliation{Department of Physics, Massachusetts Institute of Technology, Cambridge, MA 02139, USA}
\author{E.~Lopez-Asamar} \affiliation{Departamento de F\'{\i}sica Te\'orica and Instituto de F\'{\i}sica Te\'orica UAM/CSIC, Universidad Aut\'onoma de Madrid, 28049 Madrid, Spain}
\author{R.~Mahapatra} \affiliation{Department of Physics, Texas A \& M University, College Station, TX 77843, USA}
\author{V.~Mandic} \affiliation{School of Physics \& Astronomy, University of Minnesota, Minneapolis, MN 55455, USA}
\author{C.~Martinez} \affiliation{Department of Physics, Queen's University, Kingston, ON K7L 3N6, Canada}
\author{K.A.~McCarthy} \affiliation{Department of Physics, Massachusetts Institute of Technology, Cambridge, MA 02139, USA}
\author{N.~Mirabolfathi} \affiliation{Department of Physics, University of California, Berkeley, CA 94720, USA}
\author{R.A.~Moffatt} \affiliation{Department of Physics, Stanford University, Stanford, CA 94305, USA}
\author{D.C.~Moore} \affiliation{Division of Physics, Mathematics \& Astronomy, California Institute of Technology, Pasadena, CA 91125, USA}
\author{P.~Nadeau} \affiliation{Department of Physics, Queen's University, Kingston, ON K7L 3N6, Canada}
\author{R.H.~Nelson} \affiliation{Division of Physics, Mathematics \& Astronomy, California Institute of Technology, Pasadena, CA 91125, USA}
\author{K.~Page} \affiliation{Department of Physics, Queen's University, Kingston, ON K7L 3N6, Canada}
\author{R.~Partridge} \affiliation{SLAC National Accelerator Laboratory / Kavli Institute for Particle Astrophysics and Cosmology, 2575 Sand Hill Road, Menlo Park 94025, CA}
\author{M.~Pepin} \affiliation{School of Physics \& Astronomy, University of Minnesota, Minneapolis, MN 55455, USA}
\author{A.~Phipps} \affiliation{Department of Physics, University of California, Berkeley, CA 94720, USA}
\author{K.~Prasad} \affiliation{Department of Physics, Texas A \& M University, College Station, TX 77843, USA}
\author{M.~Pyle} \affiliation{Department of Physics, University of California, Berkeley, CA 94720, USA}
\author{H.~Qiu} \affiliation{Department of Physics, Southern Methodist University, Dallas, TX 75275, USA}
\author{W.~Rau} \affiliation{Department of Physics, Queen's University, Kingston, ON K7L 3N6, Canada}
\author{P.~Redl} \affiliation{Department of Physics, Stanford University, Stanford, CA 94305, USA}
\author{A.~Reisetter} \affiliation{Department of Physics, University of Evansville, Evansville, IN 47722, USA}
\author{Y.~Ricci} \affiliation{Department of Physics, Queen's University, Kingston, ON K7L 3N6, Canada}
\author{T.~Saab} \affiliation{Department of Physics, University of Florida, Gainesville, FL 32611, USA}
\author{B.~Sadoulet} \affiliation{Department of Physics, University of California, Berkeley, CA 94720, USA}\affiliation{Lawrence Berkeley National Laboratory, Berkeley, CA 94720, USA}
\author{J.~Sander} \affiliation{Department of Physics, Texas A \& M University, College Station, TX 77843, USA}
\author{K.~Schneck} \affiliation{SLAC National Accelerator Laboratory / Kavli Institute for Particle Astrophysics and Cosmology, 2575 Sand Hill Road, Menlo Park 94025, CA}
\author{R.W.~Schnee} \affiliation{Department of Physics, Syracuse University, Syracuse, NY 13244, USA}
\author{S.~Scorza} \affiliation{Department of Physics, Southern Methodist University, Dallas, TX 75275, USA}
\author{B.~Serfass} \affiliation{Department of Physics, University of California, Berkeley, CA 94720, USA}
\author{B.~Shank} \affiliation{Department of Physics, Stanford University, Stanford, CA 94305, USA}
\author{D.~Speller} \affiliation{Department of Physics, University of California, Berkeley, CA 94720, USA}
\author{K.M.~Sundqvist} \affiliation{Department of Physics, University of California, Berkeley, CA 94720, USA}
\author{A.N.~Villano} \affiliation{School of Physics \& Astronomy, University of Minnesota, Minneapolis, MN 55455, USA}
\author{B.~Welliver} \affiliation{Department of Physics, University of Florida, Gainesville, FL 32611, USA}
\author{D.H.~Wright} \affiliation{SLAC National Accelerator Laboratory / Kavli Institute for Particle Astrophysics and Cosmology, 2575 Sand Hill Road, Menlo Park 94025, CA}
\author{S.~Yellin} \affiliation{Department of Physics, Stanford University, Stanford, CA 94305, USA}
\author{J.J.~Yen} \affiliation{Department of Physics, Stanford University, Stanford, CA 94305, USA}
\author{J.~Yoo} \affiliation{Fermi National Accelerator Laboratory, Batavia, IL 60510, USA}
\author{B.A.~Young} \affiliation{Department of Physics, Santa Clara University, Santa Clara, CA 95053, USA}
\author{J.~Zhang} \affiliation{School of Physics \& Astronomy, University of Minnesota, Minneapolis, MN 55455, USA}

\collaboration{CDMS Collaboration}

\noaffiliation

\begin{abstract}
We report results of a search for Weakly Interacting Massive Particles (WIMPs) with the Si detectors of the CDMS II experiment.  This report describes a blind analysis of the first data taken with CDMS II's full complement of detectors in 2006-2007; results from this exposure using the Ge detectors have already been presented.  We observed no candidate WIMP-scattering events in an exposure of 55.9 kg-days before analysis cuts, with an expected background of $\sim$1.1 events.  The exposure of this analysis is equivalent to 10.3 kg-days over a recoil energy range of 7-100 keV for an ideal Si detector and a WIMP mass of 10 \gev.  These data set an upper limit of $1.7\times10^{-41}$ cm$^2$ on the WIMP-nucleon spin-independent cross section of a 10 \gev WIMP.  These data exclude parameter space for spin-independent WIMP-nucleon elastic scattering that is relevant to recent searches for low-mass WIMPs.
\end{abstract}


\maketitle


There is now overwhelming evidence that the bulk of the matter in our universe is in some nonluminous, nonbaryonic form \cite{Bertone:2004pz}.  Though there is broad consensus on the amount of this dark matter present in the cosmos, its composition has thus far eluded laboratory investigations.  Weakly Interacting Massive Particles (WIMPs) \cite{Steigman:1984ac} -- particles with masses between a few \gev and a few \tev and interaction strengths characteristic of the weak force -- form a leading class of candidates for this dark matter.  Particles of this type would be produced thermally in the early universe in roughly the correct amount, and are predicted by many theoretical extensions to the Standard Model of particle physics \cite{Lee:1977ua,*Jungman:1995df,*Bertone:2004pz}.  If WIMPs do constitute the dark matter in our galaxy, they may be detectable through their elastic scattering off of nuclei in terrestrial particle detectors \cite{Goodman:1984dc}.  Numerous experimental groups have sought to detect such scattering events using a wide variety of technologies \cite{Bertone:2010zza}.

The Cryogenic Dark Matter Search (CDMS) collaboration seeks to identify nuclear recoils induced by WIMP interactions using semiconductor detectors operated at very low temperatures ($\sim$40 mK).  These detectors use a simultaneous measurement of ionization and out-of-equilibrium phonons to identify such events among a far more numerous background of electron recoils.  From 2003-2008 the collaboration operated CDMS II, an array of Ge and Si detectors located at the Soudan Underground Laboratory.  Previous results from the CDMS II installation \cite{Akerib:2004fq,Akerib:2005kh,Ahmed:2008eu,CDMSScience:2010} have set stringent upper limits on the WIMP-nucleon scattering cross section and constrained some non-WIMP dark matter candidates \cite{Ahmed:2010hw,*Ahmed:2009rh,*Ahmed:2009ht}.

This work presents results from a search for WIMP interactions in the CDMS II Si detectors during the first run of the experiment with its full complement of detectors.  The lower atomic mass of Si generally makes it a less sensitive target for spin-independent (scalar) WIMP interactions, due to the coherent enhancement of the scattering cross section for heavy nuclei.  The lower atomic mass of Si is advantageous in searches for WIMPs of relatively low mass, however, due to more favorable scattering kinematics.  A WIMP of mass $\lesssim40$ \gev will impart more recoil energy to a Si atom than to a Ge atom on average, so a WIMP of sufficiently low mass ($M\lesssim10$ \gev for CDMS II) will generate more detectable recoils in a Si detector at fixed energy threshold.  New particles at such masses are generally disfavored in fits of supersymmetry models to precision electroweak data (\eg \cite{Baltz:2004aw,*Roszkowski:2007fd}), but viable models in this regime do exist (\eg \cite{Bottino:2003cz,*Kaplan:2009ag,*Cohen:2009fz}).  Renewed interest in this mass range has been motivated by results from the DAMA/LIBRA \cite{Bernabei:2008yi}, CoGeNT \cite{Aalseth:2010vx}, and CRESST \cite{Angloher:2011uu} experiments, which have been interpreted as possible evidence of WIMP scattering.  CDMS has previously explored similar parameter space using dedicated low-threshold analyses of data from its shallow and deep runs \cite{Akerib:2010pv,*Ahmed:2010wy}.

In its final configuration, the CDMS II array consisted of 30 Z-sensitive ionization and phonon (ZIP) detectors: 19 Ge ($\sim$239 g each) and 11 Si ($\sim$106 g each), for a total of $\sim$4.6 kg of Ge and $\sim$1.2 kg of Si.  Each CDMS detector is a semiconductor disk, 7.6 cm in diameter and 1 cm thick, instrumented to detect the phonons and ionization generated by particle interactions within the crystal.  One flat face of each detector is instrumented with four readout channels composed of superconducting transition-edge sensors (TESs) to detect out-of-equilibrium phonons.  The opposite flat face is divided into two concentric ionization electrodes: an inner (primary) electrode covering $\sim$85\% of the detector surface and an outer guard ring.  The latter defines a fiducial volume within each ZIP by identifying interactions near the detector rim, which may suffer from reduced ionization collection.  We discriminate nuclear recoils from background electron recoils using the ratio of ionization to phonon recoil energy (``ionization yield'').
  Electron recoils that occur within $\sim$10 $\mu$m of a detector surface can be misclassified as nuclear recoils due to reduced ionization collection.  Such surface events are identified by the faster arrival of their phonon signals, giving an overall misidentification rate for electron recoils of less than 1 in $10^6$ for recoils in the energy range of greatest interest (a few tens of keV) in either detector material.

This detector array was housed within a low-radioactivity cryogenic installation \cite{Akerib:2005zy,Ahmed:2008eu} at the Soudan Underground Laboratory, Minnesota, U.S.A.  
The rock overburden above the Soudan facility (2090 meters water equivalent) reduces the flux of cosmogenic muons incident upon the detector installation by a factor of $\sim$$10^5$, thus greatly reducing the background neutron flux.  An outer hermetic layer of plastic scintillator identifies remaining cosmogenic muons entering the passive Pb and polyethylene shielding surrounding the detector volume.

We consider data taken with the Si detectors during the first two cryogenic run periods of the full CDMS II detector installation, acquired between October 2006 and July 2007.  The Ge results from this data set were described in a previous publication \cite{Ahmed:2008eu}, which was released before the Si analysis was complete.   The full CDMS II exposure at Soudan includes four later cryogenic run periods, with broadly similar instrument performance but some variation in the performance of individual detectors.  This second collection of run periods was the subject of an independent blind analysis, the results of which are reported separately \cite{CDMSScience:2010,Agnese:2013rvf}.

Of the 11 Si detectors, five were excluded from this WIMP-search analysis: two due to wiring failures that led to incomplete collection of the ionization signal, one due to unstable response on one of its four phonon channels, and two due to inadequate rejection of calibration surface events (\ba-induced events with low ionization yield) in the analysis chosen for this exposure.  These latter two detectors were in the end positions of their respective detector stacks and so did not benefit fully from our use of adjacent detectors to tag multiple-scattered particles, a particularly useful technique for characterizing near-surface electron recoils.  Periods of poor performance on individual detectors, as identified by a series of Kolmogorov-Smirnov tests, were also excluded from analysis.  After all such exclusions, these data represent 55.9 kg-days of exposure with the remaining six Si detectors before selection of WIMP candidates.

The response of these detectors to electron recoils was calibrated using extensive (several million events) exposures to radioactive \ba sources {\it in situ} at Soudan.  Electron recoils from these sources were used to empirically characterize and correct for the dependence of phonon pulse shape on event position and energy.  As in the analysis of the Ge detectors, events at large detector radii were excluded due to degraded performance of this correction technique.  Because the Si detectors generally do not show a clear 356-keV spectral line from the \ba source, their ionization and phonon energy scales were calibrated using 356-keV events that share their energy with a neighboring detector.  After these calibrations, the recoil energy of each particle event is taken to be the calibrated phonon energy, less the contribution from phonons emitted during the drift of the collected charge carriers \cite{Neganov:1985,*luke:6858}.

The detectors' response to nuclear recoils was characterized using neutron-scattering events from {\it in situ} calibration with a \cf source ($\gtrsim$3000 nuclear recoils per detector).  The resulting nuclear recoil population was used to tune the various WIMP-selection criteria of this analysis, notably those for ionization yield and phonon timing.  We have verified the calibration of the nuclear recoil energy scale by comparisons to Monte Carlo simulations of the \cf exposures, an analysis which will be described in a separate publication \cite{CDMSNRscale:2013}.  Such comparisons are particularly robust for the Si detectors, due to a fortuitous resonant feature in the Si-neutron elastic scattering cross section that appears near 20 keV recoil energy.  This study indicates that our reconstructed energy may be 10\% lower than the true recoil energy in the relevant energy range.  This would weaken our quoted results only slightly, as described below.

\begin{figure}[tb]
\centering
\includegraphics[width=3in]{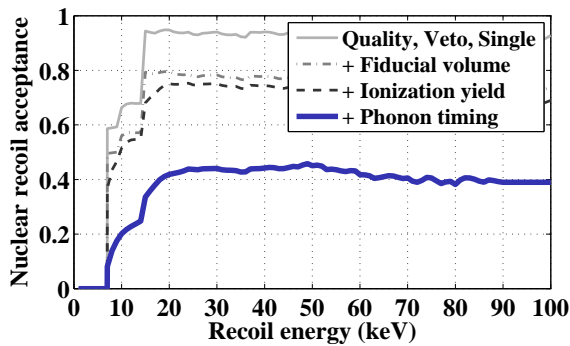}
\caption{Nuclear recoil acceptance as a function of recoil energy after successive application of each WIMP-selection criterion shown.  The bold solid curve shows the overall efficiency of this analysis.  The abrupt drops in acceptance at low recoil energies reflect the elevated energy thresholds chosen for some detectors.}
\label{fig:SiEff}
\end{figure}

\begin{figure}[tb]
\centering
\includegraphics[width=3in]{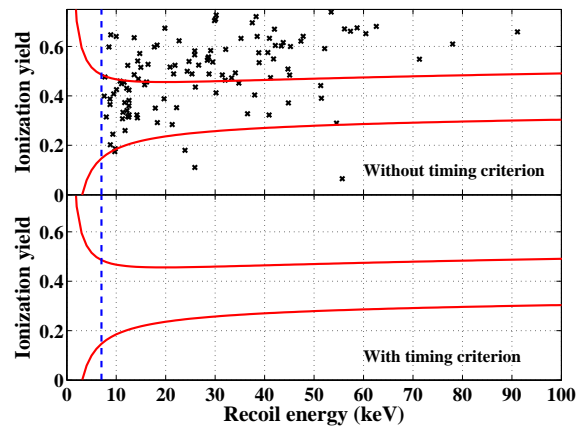}
\caption{Ionization yield versus recoil energy in all detectors included in this analysis for events passing all signal criteria except ({\it top}) and including ({\it bottom}) the phonon timing criterion.  The curved lines indicate the signal region ($\pm2\sigma$ from mean nuclear recoil yield) between 7 and 100 keV recoil energies.  Electron recoils in the detector bulk have yield near unity, above the vertical scale limits.}
\label{fig:c34SiTwoPanel}
\end{figure}

Candidate WIMP-scattering events were identified by a series of selection criteria.  These criteria were defined in parallel with those described in \cite{Ahmed:2008eu} for the Ge detectors using the same techniques.  As with the Ge detectors, all WIMP-selection criteria were defined blindly using calibration and masked WIMP-search data; for the latter, events in and near the WIMP-candidate region were automatically masked from the data set during analysis and thus had no impact on the definition of the selection criteria.  A WIMP candidate was required to have phonon and ionization signals inconsistent with noise alone, to exhibit no coincident energy in the scintillating veto shield or in any of the other 29 ZIP detectors, and not to be coincident with beam spills of the NuMI neutrino beam \cite{Anderson:1998zza}.  We further demanded that any candidate event occur within the detector's fiducial volume and have ionization yield and phonon pulse timing consistent with a nuclear recoil.  The recoil energy of each candidate event must also lie below 100 keV and above a detector-dependent threshold ranging from 7 to 15 keV.  Each detector's threshold was chosen to maintain good performance (high signal acceptance and low misidentification rate) of the phonon pulse timing criterion in calibration data, based upon the measured degradation of each detector's discrimination power at low recoil energies.  Fig.~\ref{fig:SiEff} shows the estimated fraction of WIMP-scatter events that would be accepted by these signal criteria.  Signal acceptance was measured using nuclear recoils from \cf calibration.  Monte Carlo simulations indicate that multiple-scattered neutrons in calibration data reduce the measured efficiency of the fiducial volume selection by $\sim$5.5\% with respect to the true value for single-scatter nuclear recoils, so we have scaled its efficiency upward by this amount.  Signal acceptance is $\sim$40\% at most recoil energies, somewhat higher than that of the Ge analysis.  After applying all selection criteria, the exposure of this analysis is equivalent to 10.3 kg-days over a recoil energy range of 7-100 keV for a WIMP of mass 10 \gev.

Neutrons from cosmogenic or radioactive processes can produce nuclear recoils that are indistinguishable from those from an incident WIMP.  Simulations of the rates of these processes using GEANT4 and FLUKA lead us to expect $<0.1$ false candidate events in the Si detectors from neutrons in this exposure.

A greater source of background is the misidentification of surface electron recoils, which may suffer from reduced ionization yield.  As in the Ge analysis, we developed a Bayesian estimate of the rate of misidentified surface events based upon the observed performance of the phonon timing cut for events near the WIMP-search signal region \cite{Filippini:2008}.  For the Si analysis we based our model only upon multiple-scatter events within the ionization yield acceptance region, since other event samples incorporated into the Ge analysis were found to be less reliable predictors for Si.  This model is not applicable to detectors at the top and bottom of their respective stacks, since it is impossible to identify multiple-scatter events on the outside face of such detectors.  We thus decided to exclude detectors in these positions from this blind analysis, as noted earlier.  The final model predicts an average of $1.1^{+0.9}_{-0.6} (stat.) \pm 0.1 (syst.)$ misidentified surface events in the six Si detectors during this exposure.

\begin{figure}[tb]
\centering
\includegraphics[width=3in]{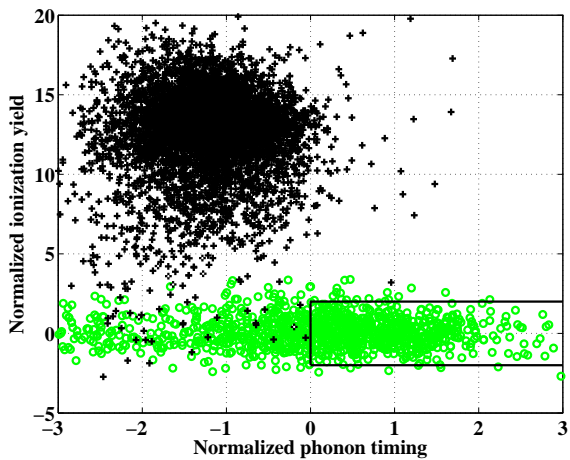}
\caption{Normalized ionization yield (standard deviations from the nuclear recoil band centroid) versus normalized phonon timing parameter ($\mu$s from the timing criterion) for events in all detectors from the WIMP-search data set passing all other selection criteria.  The black box indicates the WIMP candidate selection region.  Also plotted are nuclear recoils from \cf calibration data ({\it light, green dots}).}
\label{fig:c34SiSignalBox}
\end{figure}

After all WIMP-selection criteria were defined and the background estimate finalized, the signal regions of the Si detectors were unmasked on December 3, 2008.  No candidate WIMP-scattering events were observed.  Fig.~\ref{fig:c34SiTwoPanel} illustrates the distribution of events in and near the signal region of the WIMP-search data set before ({\it top}) and after ({\it bottom}) application of the phonon timing criterion.  Fig.~\ref{fig:c34SiSignalBox} shows an alternate view of these events, expressed in ``normalized'' versions of yield and timing that are transformed so that the WIMP acceptance regions of all detectors coincide.

\begin{figure}[tbh]
\centering
\includegraphics[width=3in]{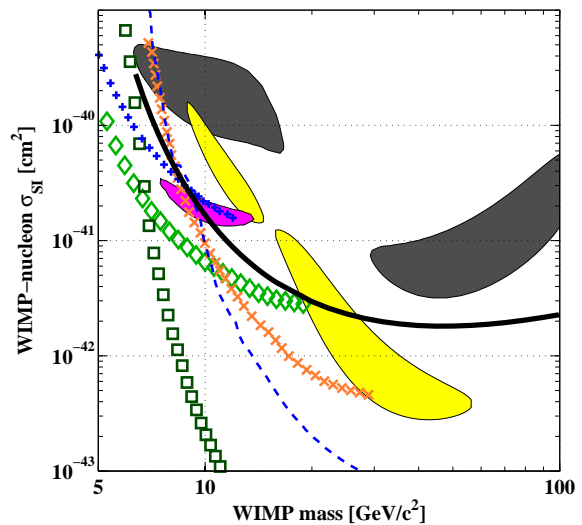}
\caption{Comparison of 90\% C.L. upper limits from these data ({\it solid}) 
with those from CDMS II Ge \cite{CDMSScience:2010,Ahmed:2010wy} ({\it dash}, +), EDELWEISS \cite{Armengaud:2012pfa} (x), XENON10 (S2-only analysis \cite{Angle:2011th,Angle:2011thErratum}, $\Diamond$), and XENON100 \cite{Aprile:2012nq} ($\Box$).  The filled regions identify regions of interest associated with data from DAMA/LIBRA \cite{Bernabei:2008yi,Savage:2008er} ({\it dark grey}, 99.7\% C.L.), CoGeNT \cite{Aalseth:2011wp} as interpreted by Kelso \etal \cite{Kelso:2011gd} ({\it magenta}, 90\% C.L., including the effect of a residual surface event contamination), and CRESST II \cite{Angloher:2011uu} ({\it yellow}, 95.4\% C.L.).}
\label{fig:SILimits}
\end{figure}

This null result constrains the available parameter space of WIMP dark matter models.  We compute upper limits on the WIMP-nucleon scattering cross section using Yellin's optimum interval method \cite{Yellin:2002xd}; this is equivalent to a Poisson upper limit in the present zero-event case, but generally results in a stronger limit when events are observed.  We work within the ``standard'' halo model described in \cite{Lewin:1995rx}, assuming a Galactic escape velocity of 544 km/s \cite{Smith:2006ym}.  Fig.~\ref{fig:SILimits} shows upper limits on the WIMP-nucleon spin-independent scattering cross section at the 90\% confidence level from CDMS II data and a selection of other recent results.  The present data set an upper limit of $1.66\times10^{-41}$ ($1.86\times10^{-42}) $ cm$^2$ for a WIMP of mass 10 (60) \gev.  The effect of a possible $\sim$10\% increase in our nuclear recoil energy scale is well approximated below 20 \gev by shifting the limit curve parallel to the mass axis by $\sim$7\%.  
Since unblinding these data, recent results from CDMS II \cite{CDMSScience:2010,Ahmed:2010wy}, EDELWEISS \cite{Armengaud:2012pfa}, XENON100 \cite{Aprile:2012nq}, and a novel low-threshold analysis of data from XENON10 \cite{Angle:2011th} also disfavor this parameter space. 

Fig.~\ref{fig:SILimits} also compares these results to three results from other instruments that have been interpreted as evidence for WIMP interactions.  The CoGeNT experiment has reported an excess of events in their Ge crystal above expected background \cite{Aalseth:2010vx} and an annual modulation of their low-energy event rate \cite{Aalseth:2011wp,Kelso:2011gd}, similar to what might be expected from interactions of a low-mass WIMP.  The CRESST II experiment has also observed an excess of events above their background model \cite{Angloher:2011uu}.  This null result disfavors portions of the best-fit regions suggested by the authors in both cases, as well as an interpretation of the DAMA/LIBRA annual modulation signal in terms of spin-independent scattering \cite{Savage:2008er}.

During the preparation of this manuscript, a similar blind analysis of the remaining CDMS II Si exposure has been completed \cite{Agnese:2013rvf}.  
That work benefits from improved analysis, calibration, and background estimation techniques that were not available for this analysis.  Additional (non-blind) studies of the combined CDMS II data set with reduced energy threshold are also planned.

The CDMS collaboration gratefully acknowledges the contributions of numerous engineers and technicians; we would like to especially thank Dennis Seitz, Jim Beaty, Bruce Hines, Larry Novak, Richard Schmitt and Astrid Tomada.  In addition, we gratefully acknowledge assistance from the staff of the Soudan Underground Laboratory and the Minnesota Department of Natural Resources. This work is supported in part 
by the National Science Foundation (Grant Nos.\ AST-9978911, NSF-1102795, PHY-0847342, PHY-0542066, PHY-0503729, PHY-0503629,  PHY-0503641, PHY-0504224, PHY-0705052, PHY-0801708, PHY-0801712, PHY-0802575, PHY-0847342, PHY-0855299, PHY-0855525, PHY-1151869, and PHY-1205898), 
by the Department of Energy (Contracts DE-AC03-76SF00098, DE-FG02-92ER40701, DE-FG03-90ER40569, and DE-FG03-91ER40618, and DE-SC0004022), 
by the Swiss National Foundation (SNF Grant No. 20-118119), 
by NSERC Canada (Grants SAPIN 341314 and SAPPJ 386399), 
and by MULTIDARK CSD2009-00064 and FPA2012-34694. 
Fermilab is operated by Fermi Research Alliance, LLC under Contract No. De-AC02-07CH11359, while SLAC is operated under Contract No.  DE-AC02-76SF00515 with the United States Department of Energy.

\bibliography{R123Si}

\begin{thebibliography}{40}%
\makeatletter
\providecommand \@ifxundefined [1]{%
 \@ifx{#1\undefined}
}%
\providecommand \@ifnum [1]{%
 \ifnum #1\expandafter \@firstoftwo
 \else \expandafter \@secondoftwo
 \fi
}%
\providecommand \@ifx [1]{%
 \ifx #1\expandafter \@firstoftwo
 \else \expandafter \@secondoftwo
 \fi
}%
\providecommand \natexlab [1]{#1}%
\providecommand \enquote  [1]{``#1''}%
\providecommand \bibnamefont  [1]{#1}%
\providecommand \bibfnamefont [1]{#1}%
\providecommand \citenamefont [1]{#1}%
\providecommand \href@noop [0]{\@secondoftwo}%
\providecommand \href [0]{\begingroup \@sanitize@url \@href}%
\providecommand \@href[1]{\@@startlink{#1}\@@href}%
\providecommand \@@href[1]{\endgroup#1\@@endlink}%
\providecommand \@sanitize@url [0]{\catcode `\\12\catcode `\$12\catcode
  `\&12\catcode `\#12\catcode `\^12\catcode `\_12\catcode `\%12\relax}%
\providecommand \@@startlink[1]{}%
\providecommand \@@endlink[0]{}%
\providecommand \url  [0]{\begingroup\@sanitize@url \@url }%
\providecommand \@url [1]{\endgroup\@href {#1}{\urlprefix }}%
\providecommand \urlprefix  [0]{URL }%
\providecommand \Eprint [0]{\href }%
\providecommand \doibase [0]{http://dx.doi.org/}%
\providecommand \selectlanguage [0]{\@gobble}%
\providecommand \bibinfo  [0]{\@secondoftwo}%
\providecommand \bibfield  [0]{\@secondoftwo}%
\providecommand \translation [1]{[#1]}%
\providecommand \BibitemOpen [0]{}%
\providecommand \bibitemStop [0]{}%
\providecommand \bibitemNoStop [0]{.\EOS\space}%
\providecommand \EOS [0]{\spacefactor3000\relax}%
\providecommand \BibitemShut  [1]{\csname bibitem#1\endcsname}%
\let\auto@bib@innerbib\@empty
\bibitem [{\citenamefont {Bertone}\ \emph {et~al.}(2005)\citenamefont
  {Bertone}, \citenamefont {Hooper},\ and\ \citenamefont
  {Silk}}]{Bertone:2004pz}%
  \BibitemOpen
  \bibfield  {author} {\bibinfo {author} {\bibfnamefont {G.}~\bibnamefont
  {Bertone}}, \bibinfo {author} {\bibfnamefont {D.}~\bibnamefont {Hooper}}, \
  and\ \bibinfo {author} {\bibfnamefont {J.}~\bibnamefont {Silk}},\ }\href
  {\doibase 10.1016/j.physrep.2004.08.031} {\bibfield  {journal} {\bibinfo
  {journal} {Phys. Rept.}\ }\textbf {\bibinfo {volume} {405}},\ \bibinfo
  {pages} {279} (\bibinfo {year} {2005})},\ \Eprint
  {http://arxiv.org/abs/hep-ph/0404175} {arXiv:hep-ph/0404175 [hep-ph]}
  \BibitemShut {NoStop}%
\bibitem [{\citenamefont {Steigman}\ and\ \citenamefont
  {Turner}(1985)}]{Steigman:1984ac}%
  \BibitemOpen
  \bibfield  {author} {\bibinfo {author} {\bibfnamefont {G.}~\bibnamefont
  {Steigman}}\ and\ \bibinfo {author} {\bibfnamefont {M.~S.}\ \bibnamefont
  {Turner}},\ }\href {\doibase 10.1016/0550-3213(85)90537-1} {\bibfield
  {journal} {\bibinfo  {journal} {Nucl. Phys.}\ }\textbf {\bibinfo {volume}
  {B253}},\ \bibinfo {pages} {375} (\bibinfo {year} {1985})}\BibitemShut
  {NoStop}%
\bibitem [{\citenamefont {Lee}\ and\ \citenamefont
  {Weinberg}(1977)}]{Lee:1977ua}%
  \BibitemOpen
  \bibfield  {author} {\bibinfo {author} {\bibfnamefont {B.~W.}\ \bibnamefont
  {Lee}}\ and\ \bibinfo {author} {\bibfnamefont {S.}~\bibnamefont {Weinberg}},\
  }\href {\doibase 10.1103/PhysRevLett.39.165} {\bibfield  {journal} {\bibinfo
  {journal} {Phys. Rev. Lett.}\ }\textbf {\bibinfo {volume} {39}},\ \bibinfo
  {pages} {165} (\bibinfo {year} {1977})}\BibitemShut {NoStop}%
\bibitem [{\citenamefont {Jungman}\ \emph {et~al.}(1996)\citenamefont
  {Jungman}, \citenamefont {Kamionkowski},\ and\ \citenamefont
  {Griest}}]{Jungman:1995df}%
  \BibitemOpen
  \bibfield  {author} {\bibinfo {author} {\bibfnamefont {G.}~\bibnamefont
  {Jungman}}, \bibinfo {author} {\bibfnamefont {M.}~\bibnamefont
  {Kamionkowski}}, \ and\ \bibinfo {author} {\bibfnamefont {K.}~\bibnamefont
  {Griest}},\ }\href {\doibase 10.1016/0370-1573(95)00058-5} {\bibfield
  {journal} {\bibinfo  {journal} {Phys. Rept.}\ }\textbf {\bibinfo {volume}
  {267}},\ \bibinfo {pages} {195} (\bibinfo {year} {1996})},\ \Eprint
  {http://arxiv.org/abs/hep-ph/9506380} {arXiv:hep-ph/9506380 [hep-ph]}
  \BibitemShut {NoStop}%
\bibitem [{\citenamefont {Goodman}\ and\ \citenamefont
  {Witten}(1985)}]{Goodman:1984dc}%
  \BibitemOpen
  \bibfield  {author} {\bibinfo {author} {\bibfnamefont {M.~W.}\ \bibnamefont
  {Goodman}}\ and\ \bibinfo {author} {\bibfnamefont {E.}~\bibnamefont
  {Witten}},\ }\href {\doibase 10.1103/PhysRevD.31.3059} {\bibfield  {journal}
  {\bibinfo  {journal} {Phys. Rev.}\ }\textbf {\bibinfo {volume} {D31}},\
  \bibinfo {pages} {3059} (\bibinfo {year} {1985})}\BibitemShut {NoStop}%
\bibitem [{\citenamefont {Bertone}(2010)}]{Bertone:2010zza}%
  \BibitemOpen
  \bibinfo {editor} {\bibfnamefont {G.}~\bibnamefont {Bertone}},\ ed.,\
  \href@noop {} {\emph {\bibinfo {title} {{Particle Dark Matter: Observations,
  Models and Searches}}}}\ (\bibinfo  {publisher} {Cambridge University
  Press},\ \bibinfo {year} {2010})\BibitemShut {NoStop}%
\bibitem [{\citenamefont {Akerib}\ \emph {et~al.}(2004)\citenamefont {Akerib}
  \emph {et~al.}}]{Akerib:2004fq}%
  \BibitemOpen
  \bibfield  {author} {\bibinfo {author} {\bibfnamefont {D.~S.}\ \bibnamefont
  {Akerib}} \emph {et~al.},\ }\href {\doibase 10.1103/PhysRevLett.93.211301}
  {\bibfield  {journal} {\bibinfo  {journal} {Phys. Rev. Lett.}\ }\textbf
  {\bibinfo {volume} {93}},\ \bibinfo {pages} {211301} (\bibinfo {year}
  {2004})},\ \Eprint {http://arxiv.org/abs/astro-ph/0405033}
  {arXiv:astro-ph/0405033 [astro-ph]} \BibitemShut {NoStop}%
\bibitem [{\citenamefont {Akerib}\ \emph {et~al.}(2006)\citenamefont {Akerib}
  \emph {et~al.}}]{Akerib:2005kh}%
  \BibitemOpen
  \bibfield  {author} {\bibinfo {author} {\bibfnamefont {D.~S.}\ \bibnamefont
  {Akerib}} \emph {et~al.},\ }\href {\doibase 10.1103/PhysRevLett.96.011302}
  {\bibfield  {journal} {\bibinfo  {journal} {Phys. Rev. Lett.}\ }\textbf
  {\bibinfo {volume} {96}},\ \bibinfo {pages} {011302} (\bibinfo {year}
  {2006})},\ \Eprint {http://arxiv.org/abs/astro-ph/0509259}
  {arXiv:astro-ph/0509259 [astro-ph]} \BibitemShut {NoStop}%
\bibitem [{\citenamefont {Ahmed}\ \emph
  {et~al.}(2009{\natexlab{a}})\citenamefont {Ahmed} \emph
  {et~al.}}]{Ahmed:2008eu}%
  \BibitemOpen
  \bibfield  {author} {\bibinfo {author} {\bibfnamefont {Z.}~\bibnamefont
  {Ahmed}} \emph {et~al.},\ }\href {\doibase 10.1103/PhysRevLett.102.011301}
  {\bibfield  {journal} {\bibinfo  {journal} {Phys. Rev. Lett.}\ }\textbf
  {\bibinfo {volume} {102}},\ \bibinfo {pages} {011301} (\bibinfo {year}
  {2009}{\natexlab{a}})},\ \Eprint {http://arxiv.org/abs/0802.3530}
  {arXiv:0802.3530 [astro-ph]} \BibitemShut {NoStop}%
\bibitem [{\citenamefont {Ahmed}\ \emph
  {et~al.}(2010{\natexlab{a}})\citenamefont {Ahmed} \emph
  {et~al.}}]{CDMSScience:2010}%
  \BibitemOpen
  \bibfield  {author} {\bibinfo {author} {\bibfnamefont {Z.}~\bibnamefont
  {Ahmed}} \emph {et~al.},\ }\href {\doibase 10.1126/science.1186112}
  {\bibfield  {journal} {\bibinfo  {journal} {Science}\ }\textbf {\bibinfo
  {volume} {327}},\ \bibinfo {pages} {1619} (\bibinfo {year}
  {2010}{\natexlab{a}})},\ \Eprint {http://arxiv.org/abs/0912.3592}
  {arXiv:0912.3592 [astro-ph.CO]} \BibitemShut {NoStop}%
\bibitem [{\citenamefont {Ahmed}\ \emph
  {et~al.}(2011{\natexlab{a}})\citenamefont {Ahmed} \emph
  {et~al.}}]{Ahmed:2010hw}%
  \BibitemOpen
  \bibfield  {author} {\bibinfo {author} {\bibfnamefont {Z.}~\bibnamefont
  {Ahmed}} \emph {et~al.},\ }\href {\doibase 10.1103/PhysRevD.83.112002}
  {\bibfield  {journal} {\bibinfo  {journal} {Phys. Rev.}\ }\textbf {\bibinfo
  {volume} {D83}},\ \bibinfo {pages} {112002} (\bibinfo {year}
  {2011}{\natexlab{a}})},\ \Eprint {http://arxiv.org/abs/1012.5078}
  {arXiv:1012.5078 [astro-ph.CO]} \BibitemShut {NoStop}%
\bibitem [{\citenamefont {Ahmed}\ \emph
  {et~al.}(2010{\natexlab{b}})\citenamefont {Ahmed} \emph
  {et~al.}}]{Ahmed:2009rh}%
  \BibitemOpen
  \bibfield  {author} {\bibinfo {author} {\bibfnamefont {Z.}~\bibnamefont
  {Ahmed}} \emph {et~al.},\ }\href {\doibase 10.1103/PhysRevD.81.042002}
  {\bibfield  {journal} {\bibinfo  {journal} {Phys. Rev.}\ }\textbf {\bibinfo
  {volume} {D81}},\ \bibinfo {pages} {042002} (\bibinfo {year}
  {2010}{\natexlab{b}})},\ \Eprint {http://arxiv.org/abs/0907.1438}
  {arXiv:0907.1438 [astro-ph.GA]} \BibitemShut {NoStop}%
\bibitem [{\citenamefont {Ahmed}\ \emph
  {et~al.}(2009{\natexlab{b}})\citenamefont {Ahmed} \emph
  {et~al.}}]{Ahmed:2009ht}%
  \BibitemOpen
  \bibfield  {author} {\bibinfo {author} {\bibfnamefont {Z.}~\bibnamefont
  {Ahmed}} \emph {et~al.},\ }\href {\doibase 10.1103/PhysRevLett.103.141802}
  {\bibfield  {journal} {\bibinfo  {journal} {Phys. Rev. Lett.}\ }\textbf
  {\bibinfo {volume} {103}},\ \bibinfo {pages} {141802} (\bibinfo {year}
  {2009}{\natexlab{b}})},\ \Eprint {http://arxiv.org/abs/0902.4693}
  {arXiv:0902.4693 [hep-ex]} \BibitemShut {NoStop}%
\bibitem [{\citenamefont {Baltz}\ and\ \citenamefont
  {Gondolo}(2004)}]{Baltz:2004aw}%
  \BibitemOpen
  \bibfield  {author} {\bibinfo {author} {\bibfnamefont {E.~A.}\ \bibnamefont
  {Baltz}}\ and\ \bibinfo {author} {\bibfnamefont {P.}~\bibnamefont
  {Gondolo}},\ }\href {\doibase 10.1088/1126-6708/2004/10/052} {\bibfield
  {journal} {\bibinfo  {journal} {JHEP}\ }\textbf {\bibinfo {volume} {10}},\
  \bibinfo {pages} {052} (\bibinfo {year} {2004})},\ \Eprint
  {http://arxiv.org/abs/hep-ph/0407039} {arXiv:hep-ph/0407039 [hep-ph]}
  \BibitemShut {NoStop}%
\bibitem [{\citenamefont {Roszkowski}\ \emph {et~al.}(2007)\citenamefont
  {Roszkowski}, \citenamefont {Ruiz~de Austri},\ and\ \citenamefont
  {Trotta}}]{Roszkowski:2007fd}%
  \BibitemOpen
  \bibfield  {author} {\bibinfo {author} {\bibfnamefont {L.}~\bibnamefont
  {Roszkowski}}, \bibinfo {author} {\bibfnamefont {R.}~\bibnamefont {Ruiz~de
  Austri}}, \ and\ \bibinfo {author} {\bibfnamefont {R.}~\bibnamefont
  {Trotta}},\ }\href {\doibase 10.1088/1126-6708/2007/07/075} {\bibfield
  {journal} {\bibinfo  {journal} {JHEP}\ }\textbf {\bibinfo {volume} {07}},\
  \bibinfo {pages} {075} (\bibinfo {year} {2007})},\ \Eprint
  {http://arxiv.org/abs/0705.2012} {arXiv:0705.2012 [hep-ph]} \BibitemShut
  {NoStop}%
\bibitem [{\citenamefont {Bottino}\ \emph {et~al.}(2004)\citenamefont
  {Bottino}, \citenamefont {Donato}, \citenamefont {Fornengo},\ and\
  \citenamefont {Scopel}}]{Bottino:2003cz}%
  \BibitemOpen
  \bibfield  {author} {\bibinfo {author} {\bibfnamefont {A.}~\bibnamefont
  {Bottino}}, \bibinfo {author} {\bibfnamefont {F.}~\bibnamefont {Donato}},
  \bibinfo {author} {\bibfnamefont {N.}~\bibnamefont {Fornengo}}, \ and\
  \bibinfo {author} {\bibfnamefont {S.}~\bibnamefont {Scopel}},\ }\href
  {\doibase 10.1103/PhysRevD.69.037302} {\bibfield  {journal} {\bibinfo
  {journal} {Phys. Rev.}\ }\textbf {\bibinfo {volume} {D69}},\ \bibinfo {pages}
  {037302} (\bibinfo {year} {2004})},\ \Eprint
  {http://arxiv.org/abs/hep-ph/0307303} {arXiv:hep-ph/0307303 [hep-ph]}
  \BibitemShut {NoStop}%
\bibitem [{\citenamefont {Kaplan}\ \emph {et~al.}(2009)\citenamefont {Kaplan},
  \citenamefont {Luty},\ and\ \citenamefont {Zurek}}]{Kaplan:2009ag}%
  \BibitemOpen
  \bibfield  {author} {\bibinfo {author} {\bibfnamefont {D.~E.}\ \bibnamefont
  {Kaplan}}, \bibinfo {author} {\bibfnamefont {M.~A.}\ \bibnamefont {Luty}}, \
  and\ \bibinfo {author} {\bibfnamefont {K.~M.}\ \bibnamefont {Zurek}},\ }\href
  {\doibase 10.1103/PhysRevD.79.115016} {\bibfield  {journal} {\bibinfo
  {journal} {Phys. Rev.}\ }\textbf {\bibinfo {volume} {D79}},\ \bibinfo {pages}
  {115016} (\bibinfo {year} {2009})},\ \Eprint {http://arxiv.org/abs/0901.4117}
  {arXiv:0901.4117 [hep-ph]} \BibitemShut {NoStop}%
\bibitem [{\citenamefont {Cohen}\ and\ \citenamefont
  {Zurek}(2010)}]{Cohen:2009fz}%
  \BibitemOpen
  \bibfield  {author} {\bibinfo {author} {\bibfnamefont {T.}~\bibnamefont
  {Cohen}}\ and\ \bibinfo {author} {\bibfnamefont {K.~M.}\ \bibnamefont
  {Zurek}},\ }\href {\doibase 10.1103/PhysRevLett.104.101301} {\bibfield
  {journal} {\bibinfo  {journal} {Phys. Rev. Lett.}\ }\textbf {\bibinfo
  {volume} {104}},\ \bibinfo {pages} {101301} (\bibinfo {year} {2010})},\
  \Eprint {http://arxiv.org/abs/0909.2035} {arXiv:0909.2035 [hep-ph]}
  \BibitemShut {NoStop}%
\bibitem [{\citenamefont {Bernabei}\ \emph {et~al.}(2008)\citenamefont
  {Bernabei} \emph {et~al.}}]{Bernabei:2008yi}%
  \BibitemOpen
  \bibfield  {author} {\bibinfo {author} {\bibfnamefont {R.}~\bibnamefont
  {Bernabei}} \emph {et~al.},\ }\href {\doibase 10.1140/epjc/s10052-008-0662-y}
  {\bibfield  {journal} {\bibinfo  {journal} {Eur. Phys. J.}\ }\textbf
  {\bibinfo {volume} {C56}},\ \bibinfo {pages} {333} (\bibinfo {year}
  {2008})},\ \Eprint {http://arxiv.org/abs/0804.2741} {arXiv:0804.2741
  [astro-ph]} \BibitemShut {NoStop}%
\bibitem [{\citenamefont {Aalseth}\ \emph
  {et~al.}(2011{\natexlab{a}})\citenamefont {Aalseth} \emph
  {et~al.}}]{Aalseth:2010vx}%
  \BibitemOpen
  \bibfield  {author} {\bibinfo {author} {\bibfnamefont {C.}~\bibnamefont
  {Aalseth}} \emph {et~al.},\ }\href {\doibase 10.1103/PhysRevLett.106.131301}
  {\bibfield  {journal} {\bibinfo  {journal} {Phys. Rev. Lett.}\ }\textbf
  {\bibinfo {volume} {106}},\ \bibinfo {pages} {131301} (\bibinfo {year}
  {2011}{\natexlab{a}})},\ \Eprint {http://arxiv.org/abs/1002.4703}
  {arXiv:1002.4703 [astro-ph.CO]} \BibitemShut {NoStop}%
\bibitem [{\citenamefont {Angloher}\ \emph {et~al.}(2012)\citenamefont
  {Angloher} \emph {et~al.}}]{Angloher:2011uu}%
  \BibitemOpen
  \bibfield  {author} {\bibinfo {author} {\bibfnamefont {G.}~\bibnamefont
  {Angloher}} \emph {et~al.},\ }\href {\doibase 10.1140/epjc/s10052-012-1971-8}
  {\bibfield  {journal} {\bibinfo  {journal} {Eur. Phys. J.}\ }\textbf
  {\bibinfo {volume} {C72}},\ \bibinfo {pages} {1971} (\bibinfo {year}
  {2012})},\ \Eprint {http://arxiv.org/abs/1109.0702} {arXiv:1109.0702
  [astro-ph.CO]} \BibitemShut {NoStop}%
\bibitem [{\citenamefont {Akerib}\ \emph {et~al.}(2010)\citenamefont {Akerib}
  \emph {et~al.}}]{Akerib:2010pv}%
  \BibitemOpen
  \bibfield  {author} {\bibinfo {author} {\bibfnamefont {D.}~\bibnamefont
  {Akerib}} \emph {et~al.},\ }\href {\doibase 10.1103/PhysRevD.82.122004}
  {\bibfield  {journal} {\bibinfo  {journal} {Phys. Rev.}\ }\textbf {\bibinfo
  {volume} {D82}},\ \bibinfo {pages} {122004} (\bibinfo {year} {2010})},\
  \Eprint {http://arxiv.org/abs/1010.4290} {arXiv:1010.4290 [astro-ph.CO]}
  \BibitemShut {NoStop}%
\bibitem [{\citenamefont {Ahmed}\ \emph
  {et~al.}(2011{\natexlab{b}})\citenamefont {Ahmed} \emph
  {et~al.}}]{Ahmed:2010wy}%
  \BibitemOpen
  \bibfield  {author} {\bibinfo {author} {\bibfnamefont {Z.}~\bibnamefont
  {Ahmed}} \emph {et~al.},\ }\href {\doibase 10.1103/PhysRevLett.106.131302}
  {\bibfield  {journal} {\bibinfo  {journal} {Phys. Rev. Lett.}\ }\textbf
  {\bibinfo {volume} {106}},\ \bibinfo {pages} {131302} (\bibinfo {year}
  {2011}{\natexlab{b}})},\ \Eprint {http://arxiv.org/abs/1011.2482}
  {arXiv:1011.2482 [astro-ph.CO]} \BibitemShut {NoStop}%
\bibitem [{\citenamefont {Akerib}\ \emph {et~al.}(2005)\citenamefont {Akerib}
  \emph {et~al.}}]{Akerib:2005zy}%
  \BibitemOpen
  \bibfield  {author} {\bibinfo {author} {\bibfnamefont {D.~S.}\ \bibnamefont
  {Akerib}} \emph {et~al.},\ }\href {\doibase 10.1103/PhysRevD.72.052009}
  {\bibfield  {journal} {\bibinfo  {journal} {Phys. Rev.}\ }\textbf {\bibinfo
  {volume} {D72}},\ \bibinfo {pages} {052009} (\bibinfo {year} {2005})},\
  \Eprint {http://arxiv.org/abs/astro-ph/0507190} {arXiv:astro-ph/0507190
  [astro-ph]} \BibitemShut {NoStop}%
\bibitem [{\citenamefont {Agnese}\ \emph {et~al.}(2013)\citenamefont {Agnese}
  \emph {et~al.}}]{Agnese:2013rvf}%
  \BibitemOpen
  \bibfield  {author} {\bibinfo {author} {\bibfnamefont {R.}~\bibnamefont
  {Agnese}} \emph {et~al.},\ }\href@noop {} {\bibfield  {journal} {\bibinfo
  {journal} {submitted to Phys. Rev. Lett.}\ } (\bibinfo {year} {2013})},\
  \Eprint {http://arxiv.org/abs/1304.4279} {arXiv:1304.4279 [hep-ex]}
  \BibitemShut {NoStop}%
\bibitem [{\citenamefont {Neganov}\ and\ \citenamefont
  {Trofimov}(1985)}]{Neganov:1985}%
  \BibitemOpen
  \bibfield  {author} {\bibinfo {author} {\bibfnamefont {B.~S.}\ \bibnamefont
  {Neganov}}\ and\ \bibinfo {author} {\bibfnamefont {V.~N.}\ \bibnamefont
  {Trofimov}},\ }\href@noop {} {\bibfield  {journal} {\bibinfo  {journal}
  {Otkrytia i izobretenia}\ }\textbf {\bibinfo {volume} {146}},\ \bibinfo
  {pages} {215} (\bibinfo {year} {1985})}\BibitemShut {NoStop}%
\bibitem [{\citenamefont {Luke}(1988)}]{luke:6858}%
  \BibitemOpen
  \bibfield  {author} {\bibinfo {author} {\bibfnamefont {P.~N.}\ \bibnamefont
  {Luke}},\ }\href {\doibase 10.1063/1.341976} {\bibfield  {journal} {\bibinfo
  {journal} {J. Appl. Phys.}\ }\textbf {\bibinfo {volume} {64}},\ \bibinfo
  {pages} {6858} (\bibinfo {year} {1988})}\BibitemShut {NoStop}%
\bibitem [{\citenamefont {{CDMS Collaboration}}()}]{CDMSNRscale:2013}%
  \BibitemOpen
  \bibfield  {author} {\bibinfo {author} {\bibnamefont {{CDMS
  Collaboration}}},\ }\href@noop {} {}\bibinfo {note} {\textit{$^{252}$Cf
  calibration analysis, in preparation}}\BibitemShut {NoStop}%
\bibitem [{\citenamefont {Anderson}\ \emph {et~al.}(1998)\citenamefont
  {Anderson} \emph {et~al.}}]{Anderson:1998zza}%
  \BibitemOpen
  \bibfield  {author} {\bibinfo {author} {\bibfnamefont {K.}~\bibnamefont
  {Anderson}} \emph {et~al.},\ }\href@noop {} {\enquote {\bibinfo {title} {{The
  NuMI Facility Technical Design Report}},}\ }\bibinfo {howpublished}
  {FERMILAB-DESIGN-1998-01} (\bibinfo {year} {1998})\BibitemShut {NoStop}%
\bibitem [{\citenamefont {Filippini}(2008)}]{Filippini:2008}%
  \BibitemOpen
  \bibfield  {author} {\bibinfo {author} {\bibfnamefont {J.~P.}\ \bibnamefont
  {Filippini}},\ }\href@noop {} {Ph.D. thesis},\ \bibinfo  {school} {University
  of California, Berkeley} (\bibinfo {year} {2008})\BibitemShut {NoStop}%
\bibitem [{\citenamefont {Armengaud}\ \emph {et~al.}(2012)\citenamefont
  {Armengaud} \emph {et~al.}}]{Armengaud:2012pfa}%
  \BibitemOpen
  \bibfield  {author} {\bibinfo {author} {\bibfnamefont {E.}~\bibnamefont
  {Armengaud}} \emph {et~al.},\ }\href {\doibase 10.1103/PhysRevD.86.051701}
  {\bibfield  {journal} {\bibinfo  {journal} {Phys. Rev.}\ }\textbf {\bibinfo
  {volume} {D86}},\ \bibinfo {pages} {051701} (\bibinfo {year} {2012})},\
  \Eprint {http://arxiv.org/abs/1207.1815} {arXiv:1207.1815 [astro-ph.CO]}
  \BibitemShut {NoStop}%
\bibitem [{\citenamefont {Angle}\ \emph {et~al.}(2011)\citenamefont {Angle}
  \emph {et~al.}}]{Angle:2011th}%
  \BibitemOpen
  \bibfield  {author} {\bibinfo {author} {\bibfnamefont {J.}~\bibnamefont
  {Angle}} \emph {et~al.},\ }\href {\doibase 10.1103/PhysRevLett.107.051301}
  {\bibfield  {journal} {\bibinfo  {journal} {Phys. Rev. Lett.}\ }\textbf
  {\bibinfo {volume} {107}},\ \bibinfo {pages} {051301} (\bibinfo {year}
  {2011})},\ \Eprint {http://arxiv.org/abs/1104.3088} {arXiv:1104.3088
  [astro-ph.CO]} \BibitemShut {NoStop}%
\bibitem [{\citenamefont {Angle}\ \emph {et~al.}(2013)\citenamefont {Angle}
  \emph {et~al.}}]{Angle:2011thErratum}%
  \BibitemOpen
  \bibfield  {author} {\bibinfo {author} {\bibfnamefont {J.}~\bibnamefont
  {Angle}} \emph {et~al.},\ }\href {\doibase 10.1103/PhysRevLett.110.249901}
  {\bibfield  {journal} {\bibinfo  {journal} {Phys. Rev. Lett.}\ }\textbf
  {\bibinfo {volume} {110}},\ \bibinfo {pages} {249901(E)} (\bibinfo {year}
  {2013})}\BibitemShut {NoStop}%
\bibitem [{\citenamefont {Aprile}\ \emph {et~al.}(2012)\citenamefont {Aprile}
  \emph {et~al.}}]{Aprile:2012nq}%
  \BibitemOpen
  \bibfield  {author} {\bibinfo {author} {\bibfnamefont {E.}~\bibnamefont
  {Aprile}} \emph {et~al.},\ }\href {\doibase 10.1103/PhysRevLett.109.181301}
  {\bibfield  {journal} {\bibinfo  {journal} {Phys. Rev. Lett.}\ }\textbf
  {\bibinfo {volume} {109}},\ \bibinfo {pages} {181301} (\bibinfo {year}
  {2012})},\ \Eprint {http://arxiv.org/abs/1207.5988} {arXiv:1207.5988
  [astro-ph.CO]} \BibitemShut {NoStop}%
\bibitem [{\citenamefont {Savage}\ \emph {et~al.}(2009)\citenamefont {Savage},
  \citenamefont {Gelmini}, \citenamefont {Gondolo},\ and\ \citenamefont
  {Freese}}]{Savage:2008er}%
  \BibitemOpen
  \bibfield  {author} {\bibinfo {author} {\bibfnamefont {C.}~\bibnamefont
  {Savage}}, \bibinfo {author} {\bibfnamefont {G.}~\bibnamefont {Gelmini}},
  \bibinfo {author} {\bibfnamefont {P.}~\bibnamefont {Gondolo}}, \ and\
  \bibinfo {author} {\bibfnamefont {K.}~\bibnamefont {Freese}},\ }\href
  {\doibase 10.1088/1475-7516/2009/04/010} {\bibfield  {journal} {\bibinfo
  {journal} {JCAP}\ }\textbf {\bibinfo {volume} {0904}},\ \bibinfo {pages}
  {010} (\bibinfo {year} {2009})},\ \Eprint {http://arxiv.org/abs/0808.3607}
  {arXiv:0808.3607 [astro-ph]} \BibitemShut {NoStop}%
\bibitem [{\citenamefont {Aalseth}\ \emph
  {et~al.}(2011{\natexlab{b}})\citenamefont {Aalseth}, \citenamefont {Barbeau},
  \citenamefont {Colaresi}, \citenamefont {Collar}, \citenamefont {Diaz~Leon}
  \emph {et~al.}}]{Aalseth:2011wp}%
  \BibitemOpen
  \bibfield  {author} {\bibinfo {author} {\bibfnamefont {C.}~\bibnamefont
  {Aalseth}}, \bibinfo {author} {\bibfnamefont {P.}~\bibnamefont {Barbeau}},
  \bibinfo {author} {\bibfnamefont {J.}~\bibnamefont {Colaresi}}, \bibinfo
  {author} {\bibfnamefont {J.}~\bibnamefont {Collar}}, \bibinfo {author}
  {\bibfnamefont {J.}~\bibnamefont {Diaz~Leon}},  \emph {et~al.},\ }\href
  {\doibase 10.1103/PhysRevLett.107.141301} {\bibfield  {journal} {\bibinfo
  {journal} {Phys. Rev. Lett.}\ }\textbf {\bibinfo {volume} {107}},\ \bibinfo
  {pages} {141301} (\bibinfo {year} {2011}{\natexlab{b}})},\ \Eprint
  {http://arxiv.org/abs/1106.0650} {arXiv:1106.0650 [astro-ph.CO]} \BibitemShut
  {NoStop}%
\bibitem [{\citenamefont {Kelso}\ \emph {et~al.}(2012)\citenamefont {Kelso},
  \citenamefont {Hooper},\ and\ \citenamefont {Buckley}}]{Kelso:2011gd}%
  \BibitemOpen
  \bibfield  {author} {\bibinfo {author} {\bibfnamefont {C.}~\bibnamefont
  {Kelso}}, \bibinfo {author} {\bibfnamefont {D.}~\bibnamefont {Hooper}}, \
  and\ \bibinfo {author} {\bibfnamefont {M.~R.}\ \bibnamefont {Buckley}},\
  }\href {\doibase 10.1103/PhysRevD.85.043515} {\bibfield  {journal} {\bibinfo
  {journal} {Phys. Rev.}\ }\textbf {\bibinfo {volume} {D85}},\ \bibinfo {pages}
  {043515} (\bibinfo {year} {2012})},\ \Eprint {http://arxiv.org/abs/1110.5338}
  {arXiv:1110.5338 [astro-ph.CO]} \BibitemShut {NoStop}%
\bibitem [{\citenamefont {Yellin}(2002)}]{Yellin:2002xd}%
  \BibitemOpen
  \bibfield  {author} {\bibinfo {author} {\bibfnamefont {S.}~\bibnamefont
  {Yellin}},\ }\href {\doibase 10.1103/PhysRevD.66.032005} {\bibfield
  {journal} {\bibinfo  {journal} {Phys. Rev.}\ }\textbf {\bibinfo {volume}
  {D66}},\ \bibinfo {pages} {032005} (\bibinfo {year} {2002})},\ \Eprint
  {http://arxiv.org/abs/physics/0203002} {arXiv:physics/0203002 [physics]}
  \BibitemShut {NoStop}%
\bibitem [{\citenamefont {Lewin}\ and\ \citenamefont
  {Smith}(1996)}]{Lewin:1995rx}%
  \BibitemOpen
  \bibfield  {author} {\bibinfo {author} {\bibfnamefont {J.~D.}\ \bibnamefont
  {Lewin}}\ and\ \bibinfo {author} {\bibfnamefont {P.~F.}\ \bibnamefont
  {Smith}},\ }\href {\doibase 10.1016/S0927-6505(96)00047-3} {\bibfield
  {journal} {\bibinfo  {journal} {Astropart. Phys.}\ }\textbf {\bibinfo
  {volume} {6}},\ \bibinfo {pages} {87} (\bibinfo {year} {1996})}\BibitemShut
  {NoStop}%
\bibitem [{\citenamefont {Smith}\ \emph {et~al.}(2007)\citenamefont {Smith},
  \citenamefont {Ruchti}, \citenamefont {Helmi}, \citenamefont {Wyse},
  \citenamefont {Fulbright} \emph {et~al.}}]{Smith:2006ym}%
  \BibitemOpen
  \bibfield  {author} {\bibinfo {author} {\bibfnamefont {M.~C.}\ \bibnamefont
  {Smith}}, \bibinfo {author} {\bibfnamefont {G.}~\bibnamefont {Ruchti}},
  \bibinfo {author} {\bibfnamefont {A.}~\bibnamefont {Helmi}}, \bibinfo
  {author} {\bibfnamefont {R.}~\bibnamefont {Wyse}}, \bibinfo {author}
  {\bibfnamefont {J.}~\bibnamefont {Fulbright}},  \emph {et~al.},\ }\href
  {\doibase 10.1111/j.1365-2966.2007.11964.x} {\bibfield  {journal} {\bibinfo
  {journal} {Mon. Not. Roy. Astron. Soc.}\ }\textbf {\bibinfo {volume} {379}},\
  \bibinfo {pages} {755} (\bibinfo {year} {2007})},\ \Eprint
  {http://arxiv.org/abs/astro-ph/0611671} {arXiv:astro-ph/0611671 [astro-ph]}
  \BibitemShut {NoStop}%
\end{thebibliography}%
\bibliographystyle{apsrev4-1}

\end{document}